\documentclass[twocolumn,showpacs,aps,preprintnumbers,amsmath,amssymb,superscriptaddress,nofootinbib]{revtex4}

\usepackage{graphicx,subfigure,multirow}
 \usepackage{longtable}
\usepackage{dcolumn}
\usepackage{bm}
\usepackage{axodraw4j}
\usepackage[all]{xy}
\usepackage{color}
\usepackage[usenames,dvipsnames]{xcolor}
\usepackage[unicode=true,pdfusetitle,
 bookmarks=true,bookmarksnumbered=false,bookmarksopen=false,citecolor=Turquoise,
 breaklinks=false,pdfborder={0 0 1},backref=false,colorlinks=true,pdfpagemode=FullScreen]
 {hyperref}

\newcommand{\gev}{\textrm{ GeV}} 
\newcommand{\tev}{\textrm{ TeV}}

\makeatletter

\newcommand{\fmslash}[2][0mu]{%
  \mathchoice
    {\fmsl@sh\displaystyle{#1}{#2}}%
    {\fmsl@sh\textstyle{#1}{#2}}%
    {\fmsl@sh\scriptstyle{#1}{#2}}%
    {\fmsl@sh\scriptscriptstyle{#1}{#2}}}
\newcommand{\fmsl@sh}[3]{%
  \m@th\ooalign{$\hfil#1\mkern#2/\hfil$\crcr$#1#3$}}
\makeatother

\newcommand{\lsim}{{\;\raise0.3ex\hbox{$<$\kern-0.75em\raise-1.1ex\hbox{$\sim$}}\;}}
\newcommand{\gsim}{{\;\raise0.3ex\hbox{$>$\kern-0.75em\raise-1.1ex\hbox{$\sim$}}\;}}

\newcommand{\beq}{\vspace{-0.2cm}\begin{equation}}
\newcommand{\eeq}{\vspace{-0.2cm}\end{equation}}
\newcommand{\bea}{\vspace{-0.2cm}\begin{eqnarray}}
\newcommand{\eea}{\vspace{-0.2cm}\end{eqnarray}}
\mathchardef\minus="002D

\usepackage{color}

\begin{document}
\title{A simple, yet subtle ``invariance'' of 
%
%
two-body decay kinematics}

\author{Kaustubh Agashe}
\affiliation{Maryland Center for Fundamental Physics, Department of Physics, University of Maryland, College Park, MD 20742, USA}
\author{Roberto Franceschini}
\affiliation{Maryland Center for Fundamental Physics, Department of Physics, University of Maryland, College Park, MD 20742, USA}
\author{Doojin Kim}
\affiliation{Maryland Center for Fundamental Physics, Department of Physics, University of Maryland, College Park, MD 20742, USA}
\preprint{UMD-PP-012-020}
\date{\today}

\begin{abstract}
We study the two-body decay of a mother particle into a
massless daughter.
We further assume that  the mother particle is {\em un}polarized
and has a
{\em generic}
boost distribution
in the laboratory frame.
In this case, we show analytically that
the laboratory frame energy distribution of the massless decay product has a {\it peak}, whose location 
is identical to the (fixed) energy of that particle in the {\it rest} frame of the corresponding mother particle. 
Given its simplicity and ``invariance''  under 
%
%
changes in
the boost distribution of the mother particle,
our finding 
should be useful for 
the determination of masses of mother particles.
In particular, we anticipate that such a procedure will then 
 {\em not} require a full reconstruction of this two-body decay chain (or for that matter,
information about the rest of the event). 
With this eventual goal in mind, 
we make a proposal for extracting the peak position by fitting the data to a well-motivated analytic function describing the shape of such an energy distribution. 
This fitting function is then tested on the 
theoretical prediction for top quark pair production and its decay, and it is found to be quite successful in this regard. 
As a proof of principle of the usefulness of our observation,
we  apply it for measuring the mass of the top quark at the LHC, using simulated data and including experimental effects.
\end{abstract}





\maketitle


It is very well-known that in the {\it rest} frame of a mother particle undergoing a two-body decay, the energy of each of the daughter particles is fixed 
in terms of mother and the daughter particle masses. 
Turning this fact around, we can determine the mass of the mother particle if we can measure these rest-frame energies of the daughter particles. 

However, 
%
%
often
the mother particle is produced in the laboratory with a boost, 
that too with a magnitude and direction which is (a priori) not known. 
%
%
%
Moreover, 
the boost of mother particles produced at hadron colliders is different in each event.
%
%
Such a boost {\em distribution} depends on the production mechanism of the particle and on the structure functions of the hadrons in the initial state of the collision, and is thus a complicated function. 
%
In turn, the fact that the mother has a different boost in each event implies that 
when we consider the {\it observed} energy of the two-body decay product in the laboratory frame, we get a {\it distribution} in it. 
Thus it seems like the information that was encoded in the rest frame energy is lost, and we are prevented from extracting (at least at an easily tractable level) the mass of the mother particle
along the lines described above.

We show that, remarkably, if one of the daughter particles from the two-body decay is massless and the mother is unpolarized, 
then such is {\em not} the case.
%
%
Specifically, in this case, 
we demonstrate that the distribution of the daughter particle's energy in the laboratory frame has a {\it peak} precisely at its corresponding rest-frame energy.

This result is interesting {\it per se}. Furthermore, 
we expect that it will lead to  
%
%
%
formulation of new methods for mass measurements. Obviously, for this purpose, 
%
%
we need to be able to determine the location of this peak accurately from the observed energy distribution of the massless daughter. To this end, we propose and motivate an analytic function that can be used to fit the data
on the energy distribution and thus extract the peak position. 
We show that this function is a suitable one
%
%
using the top quark decay, $t \to W^{-} b$, as a test case, namely, it fits very well the {\em theory} prediction for energy spectrum of the resulting $b$-jets.
Simulating a realistic  experimental situation, we then show that we can 
extract the value of the top mass from the position of the peak in the $b$-jet energy distribution along with the well-measured mass of the $W$ boson.


Let us consider the decay of a heavy particle $B$ of mass $m_{B}$, i.e., $B\rightarrow A\,a\,$
where  $a$ is a {\it massless} visible particle. For the subsequent arguments,  the properties of the particle $A$
(other than its mass denoted by $m_A$) are irrelevant.
In the rest frame of particle $B$, the energy of the particle $a$ is simply given by
\beq
E^*=\frac{m_B^2-m_A^2}{2m_B}\,. \label{Estar}
\eeq
Here and henceforth the starred quantity denotes that it is measured in the rest frame of particle $B$, i.e., the mother particle. If the mother particle (originally at rest) is  boosted by a Lorentz factor $\gamma$ 
in going to the laboratory frame, then the energy of particle $a$ seen in the laboratory frame is
\bea
E=E^*\gamma\left(1+\beta\cos\theta^*\right)\,,
\eea 
where $\theta^*$ defines the direction of emission of particle $a$ in the rest frame of $B$ with respect to the boost direction 
$\vec{\beta}$ of the mother $B$ in the laboratory frame. 
Note that both $\cos \theta^*$ and $\gamma$ can vary event-by-event. Therefore we get a {\em probability distribution} for the observed energy, which is the focus of our paper. 
Due to our assumption of the mother being not polarized,
the probability distribution of  $\cos\theta^*$ is  flat. This implies that, for a fixed $\gamma$,  the distribution of $E$ is flat as well.
More precisely, since $\cos\theta^* \in [-1,1]$, for any fixed $\gamma$ the shape of the distribution of  $E$ is a simple ``rectangle'' spanning the range~\footnote{This result is quite well-known and was used for a measurement of the $W$ mass at lepton colliders \cite{Abbiendi:2003rt}. } 
\bea
x\equiv \frac{E}{E^*}\in \left[\left(\gamma-\sqrt{\gamma^2-1}\right),\left(\gamma+\sqrt{\gamma^2-1}\right)\right], \label{eq:rangeofE} 
\eea
where $x$ defines the dimensionless energy variable of the visible particle in the laboratory frame normalized by its rest-frame energy.
%
%
A few crucial observations are in order. First, the lower (upper) bound of eq.~(\ref{eq:rangeofE}) is smaller (larger) than 1
for an arbitrary $\gamma$,
which implies that 
%
%
{\it every} rectangle 
contains $E^*$. 
%
%
Remarkably, $E^{*}$ is the only value of the energy to enjoy such a property
as long as the distribution of mother particle boost is non-vanishing in a small region around
$\gamma = 1$.
Furthermore, the energy distribution being flat for every $\gamma$, there is no other value of the energy which gets a larger contribution 
than $E^*$.
Thus, up on ``stacking up" the rectangles of different widths, corresponding to a {\em range} of $\gamma$'s, 
we see that the peak of the energy distribution of the particle $a$ is unambiguously located at $E=E^*$.
In fact, this argument goes through even for a {\em massive} daughter, provided we restrict boosts
of the mother particle to $\gamma < \left( 2 \gamma^{ * \; 2 } - 1 \right)$, where
$\gamma^*$ denotes the boost of the daughter in the rest frame of the mother.
Secondly, such rectangles are {\it asymmetric} with respect to  the point  
$E=E^*$
%
%
i.e., the upper bound is farther from 
it
%
%
than is the lower bound. Thus, 
the energy distribution of the particle $a$ has a longer tail toward high energy with respect to such a peak. 

More formally, the normalized differential decay width in $x$ for a {\em fixed} $\gamma$ is given by 
%

\begin{footnotesize}
\bea
\left.\frac{1}{\Gamma}\frac{d\Gamma}{dx}\right|_{\textnormal{fixed }\gamma}=\frac{\Theta\left(x-\gamma+\sqrt{\gamma^2-1} \right)\Theta\left( - x+\gamma+\sqrt{\gamma^2-1} \right)}{2\sqrt{\gamma^2-1}}
\label{fixedgamma}
\eea 
\end{footnotesize}
\noindent
where $\Theta(x)$ is the usual Heaviside step function, and the two step functions here merely define the allowed range of $x$.
%
%
Next, consider a probablility distribution of boosts of the mother given by $g( \gamma)$.
A given energy of daughter in laboratory frame ($x$)
can actually result from
a specific range of values of the mother boost ($\gamma$), as per eq.~(\ref{fixedgamma}).
So, we have to superpose these contributions weighted by the boost distribution, giving:
\bea
f(x)\equiv\frac{1}{\Gamma}\frac{d\Gamma}{dx}=\int_{\frac{1}{2}\left(x+\frac{1}{x} \right)}^{\infty}d\gamma \frac{g(\gamma)}{2\sqrt{\gamma^2-1}}\label{eq:f} \,.
\eea
%
%
The lower end in the integral here was derived from the solution to the equation, $x=\gamma\pm\sqrt{\gamma^2-1}$, for $\gamma$ 
where the positive (negative) signature is relevant for $x\geq 1$ $(x<1)$.
From eq.~(\ref{eq:f}) we can also compute the first derivative of $f(x)$, that is 
\bea
f'(x)=\frac{\textnormal{sgn}(1-x)}{2x}g\left(\frac{1}{2}\left(x+\frac{1}{x} \right) \right)\label{eq:fp}\,.
\eea

We assume that $g(\gamma)$ has no zeros in the range of  $\gamma$ strictly
between 1 and the kinematical limit of the collider at hand.
In what follows we show that this is sufficient to guarantee that there is a peak at $E^{*}$.  
We consider the two possibilities for $g(1)$.
Namely, if it vanishes, then $f'(x=1)\propto g(1)=0$, and the distribution has its unique extremum at $E=E^*$, following from our assumption on $g(\gamma)$. If $g(1)\neq0$, then $f'(x)$ flips its sign at $x=1$ so that the energy distribution has a cusp at $E=E^*$.
Also, 
the function $f(x)$ is positive and vanishes for both $x\to 0$ and $x\to \infty$
since those two limits lead to a trivial definite integral in eq.~(\ref{eq:f}). 
%
%
Combining all these features, we see that the point $E=E^{*}$ is necessarily the peak of the distribution 
for both values of $g(1)$.
%
%
This completes the formal proof of the peak location
for a generic boost distribution of the mother particle.

Before we move on to an application of the above result, let us pause to mention that the above ``energy-peak" is not to be confused with 
the well-known ``Jacobian peak'' which can arise in a somewhat similar situation, namely, the distribution of momentum transverse to the initial beam direction ($p_T$)
of a massless daughter from a two-body decay has a peak (under certain circumstances) at a value given by the energy of the daughter in the rest frame of the mother \cite{Han:2005mu}.
To begin with,  
%
%
 the distributions in consideration themselves are different: $p_T$ for the Jacobian peak 
vs. energy studied here.
Moreover, the 
Jacobian peak is strictly speaking applicable for a mother being boosted in longitudinal direction only,
whereas the energy-peak holds, as we discussed above, for a generic boost of the mother.
Furthermore, the events corresponding to the two peaks are distinct, in particular, 
the events at the energy-peak correspond to emission of the daughter particle in a direction (in the rest frame
of the mother particle) which has a backward
component relative to the mother boost, i.e., $\cos \theta^* = - \sqrt{ \frac{ \gamma - 1 }{ \gamma + 1 } } < 0$, as opposed to the emission being
exactly normal to the mother boost for the Jacobian peak \cite{Han:2005mu}.
%
%
Clearly, the events at the Jacobian peak can  have significant longitudinal momentum originating
from the boost of the mother so that the (total) energy in these events is larger than the
value in the rest frame, that is to say larger than for events at the energy-peak.
Note that the peak in the $p_T$ distribution is also the endpoint, a feature which is related to its Jacobian nature, whereas the energy-peak is not the endpoint.
Finally, the Jacobian peak is valid even for a 
polarized mother, as opposed to the requirement of mother being unpolarized
for the energy-peak.
Thus, there is a certain degree of complementarity in the applicability of these two peaks.

As advertised at the beginning, our finding can be utilized to measure a combination of $m_{A}$ and $m_{B}$ given by $E^{*}$, which means that $m_B$ can be 
determined
if  $m_A$ is known, or vice versa. For this purpose, we are required to extract the location of the peak accurately from data. Clearly, having a theoretical prediction from first principles for the shape of $f(x)$ is very hard because the boost distribution $g(\gamma)$ is inherently process-dependent.
%
Nevertheless, we 
know some properties of 
%
%
$f(x)$ which are listed below: i) the value of $f(x)$ remains the same under $x\leftrightarrow \frac{1}{x}$, ii) $f$ is maximized at $x=1$, iii) $f$ vanishes as $x$ approaches $0$ or $\infty$, iv)  $f$ becomes a $\delta$-function in some limit of its parameters.
The first property follows from 
the $x$-dependence of $f$ arising only
from the lower limit of the integral in
eq.~(\ref{eq:f}), and the second from eq.~(\ref{eq:fp}) and the argument thereafter. 
The third one is also manifest from eq.~(\ref{eq:f})  as mentioned above.
%
%
Finally, the last one reflects the fact that 
when the mother particle is not boosted, we get a
fixed value of energy given in eq.~(\ref{Estar}), i.e., a delta-function.
%
%

Being aware of the constraints given above, we propose the following ``simple'' function as an {\it ansatz} for $f(x)$:
\bea
f(x)=
%
%
K^{ -1 }_1(p)\exp \left[-\frac{p}{2} \left(x+\frac{1}{x} \right) \right]\,, 
\label{eq:fitter}
\eea
%
%
where  $p$ is a parameter which encodes the width of the peak and the normalization factor $K_{1}(p)$ is
a modified Bessel function of the second kind of order 1.
%
%
One can easily prove that the proposed ansatz can be reduced to a $\delta$-function for any sufficiently large $p$ using the asymptotic behavior of $K_1(p)$ such that
\bea
K_1(p) \stackrel{p\rightarrow\infty}{\longrightarrow}\; \sim \frac{e^{-p}}{\sqrt{p}}\left( 1+\mathcal{O}\left(\frac{1}{p} \right)\right).
\eea
%
%
Finally, we can show that the above ansatz does not have a cusp (at $E^*$) 
so that it is more suitable for the case of $g(1)=0$ such as pair-production of mothers.


\begin{figure}[t]
\centering
\includegraphics[width=0.95 \linewidth]{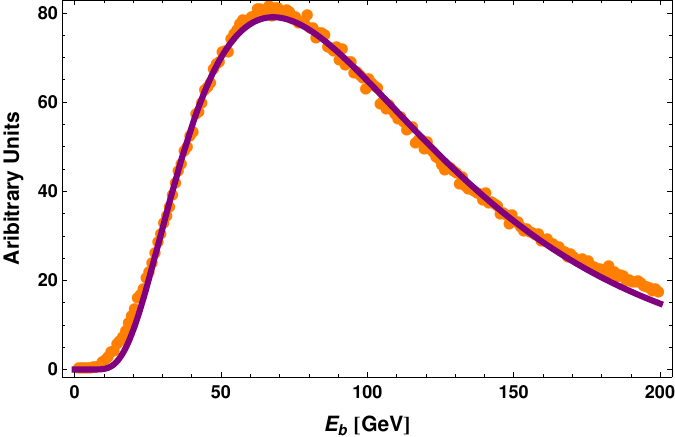}
\caption{The orange dots are the theory prediction for $d\sigma/dE_{b}$ 
in the process $pp \rightarrow t\bar{t} \to b \bar{b} \mu^{-}e^{+} \nu_{e}\bar{\nu}_{\mu}$ computed with $\mathtt{MadGraph5}$ 
at LHC with $\sqrt{s}$=7~TeV. The purple line is the best fit of our {\it ansatz} eq.~(\ref{eq:fitter}). \label{fig:ansatzcheck}}\vspace{-0.5cm}
\end{figure}

In order to test the goodness of the {\it ansatz} given in eq.~(\ref{eq:fitter}) we use it to fit a {\em theoretical} prediction for the distribution of $b$-jet energy in top decay. 
%
%
The 
bottom quark is not massless; it is nonetheless highly boosted in the rest frame of top quark, namely, $\gamma^* \approx 15$.
Based on our earlier discussion of the massive case, our argument for the peak in $b$-jet energy being at $E^*$ is invalidated
for boosts of the top quark which are so large ($\gamma \gtrsim 500$) as to have a negligible probability.
Hence, we expect the peak to be very close to $E^*$. Similarly, we expect that the first of the functional properties of the energy spectrum eq.~(\ref{eq:f}) will be only negligibly violated by the non-zero mass of the bottom quark. This justifies the use of the {\it ansatz} eq.~(\ref{eq:fitter}) to fit the $b$-jet energy spectrum.
%
%

Specifically, we study a sample of fully leptonic top decays from the process, $pp \rightarrow t\bar{t} \to b \bar{b} \mu^{-}e^{+} \nu_{e}\bar{\nu}_{\mu}$,
at the Large Hardron Collider (LHC) with 7 TeV center-of-mass energy.
%
%
To compute the theory prediction for the given process 
we employ $\mathtt{MadGraph5}$~1.4.2 \cite{Alwall:2011fk} taking $m_{\mathrm{top}}$ of 173~GeV and the patron distribution functions (PDFs) $\mathtt{CTEQ6L1}$~\cite{Pumplin:2002vw} 
with default choice of the renormalization and factorization scales.



The result of the associated fit is exhibited in FIG.~\ref{fig:ansatzcheck} which shows a very good agreement between the theory prediction from $\mathtt{MadGraph5}$ and the fitting function.  To quantify the goodness of the {\it ansatz} with an objective measure we compute both the Kolmogorov-Smirnov (KS)~\cite{James:2006zz} and the $\chi^{2}$ value. The latter is computed taking bin counts for a luminosity of $5fb^{-1}$ at LHC with $\sqrt{s}=7 \tev$ assuming that the error on each bin count is gaussian. The result is $\chi^{2}=39.3$ for 198 degrees of freedom while the KS test statistic is 0.012, which, rather than being taken in any statistical sense, should be taken as an indication that our {\it ansatz} gives a very good fit to the theory curve.
%
%
%
%
We have investigated the sensitivity of this result to the choice of the PDFs by repeating the same fit 
for the theory prediction obtained using the $\mathtt{MRST2002NLO}$ PDFs set of Ref.~\cite{Martin:2003sf}. We observe negligible differences from the result obtained with $\mathtt{CTEQ6L1}$.


So far, we have found that the {\it ansatz} in eq.~(\ref{eq:fitter}) is  very good at reproducing the theory prediction. In fact,
%
%
this success
suggests that the {\it ansatz} may be used to measure the combination of masses in eq.~(\ref{Estar}) from 
{\em experimental} data. 
%
%
In order to investigate
this possibility, we go back to the example 
of the top quark, namely, we would like to use the fitting function in order to extract the 
peak of the observed energy distribution of the $b$-jet and measure the top quark mass by plugging this value and the 
well-known mass of the $W$ boson 
into eq.~(\ref{Estar}).

Before getting into details, we would like to mention that
%
%
%
we do not necessarily aim at getting a result for the value of $m_{\mathrm{top}}$
that is competitive with the current measurements. Rather, we aim at finding what is the sensitivity of our method for
measuring top quark mass
in a realistic setup. 
In fact, for a fair comparison it should be remarked that the current measurements of $m_{\mathrm{top}}$ rely on rather complicated tools and 
often advocate templates for the distributions that require a detailed knowledge of the underlying dynamics of the top quark decay. 
On the contrary, our method is extremely simple: it is based on pure kinematics and does not rely at all on detailed knowledge of 
the above-mentioned dynamics (as long as the top quark is produced unpolarized). 
As such we can regard our study of the mass measurement of the top as a proof of principle that our method can be used
to measure the mass of heavy particles, in particular, new physics particles. 

\begin{figure}[t]
\centering
\includegraphics[width=0.95 \linewidth]{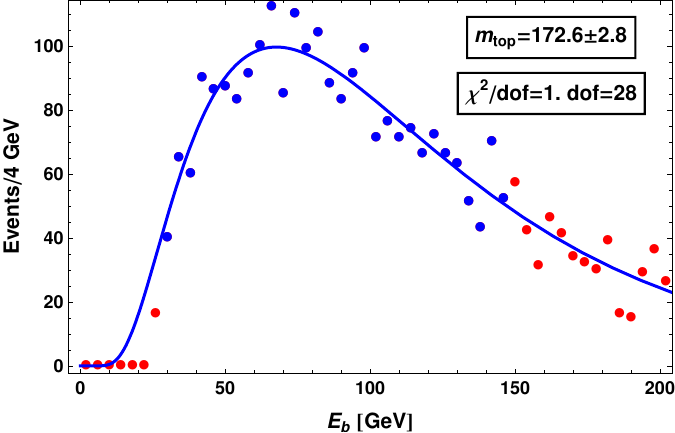}
\caption{An instance of the result of the fit on the energy distribution of the $b$-jets in a pseudo-experiment. For the fit we binned data in bins of 4 GeV. Only the blue data points are used in the fit, which correspond to using only the part of the spectrum from 30 to 150 GeV.}\label{fig:samplefit}\vspace{-0.5cm}
\end{figure}

In lieu of actual experimental data 
we use  a sample of Monte Carlo (MC)-{\em simulated} collision events.
Namely, 
we further process the previous parton-level event sample to include the effects of showering and hadronization as described in $\mathtt{PYTHIA}$~6.4~\cite{Sjostrand:2006zr} with detector response simulated by $\mathtt{Delphes}$~1.9~\cite{Ovyn:2009ys} and jets made with $\mathtt{FastJet}$~\cite{Cacciari:2011rt,Cacciari:2006vn} using the $anti$-$k_T$ algorithm \cite{Cacciari:2008hb} with the parameter choice $R=0.4$. Furthermore, we impose cuts on the final state 
following the selections of Ref.~\cite{ATLAS-CONF-2012-097} for the $e\mu$ final state.  We consider an ensemble of 100 pseudo-experiments, each of which is equivalent to $5\,fb^{-1}$ of data from the LHC at $\sqrt{s}=$~7 TeV. For each pseudo-experiment we perform a fit with our {\it ansatz} eq.~(\ref{eq:fitter}). From the extracted value of the peak of the distribution we
get a measurement of $m_{\mathrm{top}}$. The distribution over the 100 pseudo-experiments of $m_{\mathrm{top}}$ and its $1\sigma$ error are symmetric around the central values, and do not show special features. For a bin size of 4 GeV the average best-fit $m_{\mathrm{top}}$ and the $1\sigma$ error resulting from the fit are 
\begin{equation}
\langle m_{\mathrm{top}}\rangle=173.1\pm 2.5 \gev \label{mtop}
\end{equation}
with a median of the reduced $\chi^{2}$ of our fit equal to 1.1. We have checked that the result of the fit is stable under changes of the range of energies taken in the fit and changes of the bin size. For illustration purpose we show the outcome of one of the pseudo-experiments in FIG.~\ref{fig:samplefit}.

The obtained value of $m_{\mathrm{top}}$ is rather good: the error is small and the central value is compatible with the input value
within the $1\sigma$ error. 
%
%
\footnote{In our analysis there is no assessment of systematic errors and the quoted measurement error is purely statistical. The obtained statistical error projected to 300 fb$^{-1}$ at LHC  $\sqrt{s} = 14$~TeV will reduce to $\sim 200$ MeV
at LHC at $\sqrt{s} = 14$ TeV.}
Furthermore, the obtained $\chi^{2}$ is good. All this tells that the 
%
%
usefulness 
of our function eq.~(\ref{eq:fitter}) 
%
%
is not spoiled by the selection criteria
%
%
for top quark decay events nor by detector effects.
%
Even though our assessment of the power of the technique does not take the background and the entire realm of detector effects into consideration, we regard it as rather encouraging for the determination of the masses of heavy particles at colliders based only on the kinematics of the decay.  

Note that our analysis does not include resolvable -- and thus necessarily hard -- radiation
from the bottom quark in the above process. This extra radiation will turn the decay into a genuine three-body
one so that our formalism will not apply to it. Therefore, if one wants to interpret our result in eq.~(\ref{mtop}) as a realistic mass measurement, then the corrections from this process should be taken into account. Nevertheless, we have estimated that the cross-section for it is
smaller than the leading order one by one order of magnitude or more. Also, 
it might be possible to veto the extra radiation to further suppress such a contribution.

In conclusion, we have shown that for the two-body decay of an unpolarized boosted mother particle, the energy spectrum of a massless daughter in the laboratory frame encodes in a rather simple manner information about the masses involved in the decay. We showed as well how this can be used for mass measurements at hadronic colliders, which represents a remarkable twist in 
this
%
%
paradigm.
Indeed,
%
%
%
instead of using longitudinally or fully Lorentz invariant 
quantities
%
%
for this purpose, we 
extracted the mass of the top quark from a Lorentz-variant observable, i.e., the energy of the $b$-jet.
The crucial point is that even though the distribution of this quantity depends on the possible boosts of the mother particle, the peak position in it is ``invariant".
%
%
%
Another merit of our method is that 
it does not rely on any measurement of the other particle of the two-body decay 
so that we can extract some information about masses even if the latter is invisible. 
In this case, a full reconstruction of the decay chain might not be possible so that mass measurement would  not be possible with, for example, an invariant mass measurement.\footnote{ For such ``semi-invisible" decays, several techniques, based on the underlying kinematics of such processes, have been suggested and used {\em recently} for effectively carrying out the relevant {\em searches}: for example, the variables $M_{T2}$ and its variations~\cite{Lester:1999et, Barr:2003fj, Cho:2007qv, Barr:2011ao,Cho:2014naa},
$\alpha_T$~\cite{Randall:2008rw}, or razor ~\cite{Rogan:2010kb} (for a review, 
see, for example, Ref.~\cite{Gripaios:2011kk}).
{\em Post}-discovery of such new particles, the focus will shift to mass {\em measurement}, where 
the $M_{ T2}$ variable (and related 
%
%
ones) \cite{Lester:1999et, Barr:2003fj, Cho:2007qv, Barr:2011ao,Cho:2014naa}
have been designed to determine the individual new particle masses, making use of missing transverse momentum
(MET) originating from the invisible particle. See also, for example,
Refs.~\cite{Cheng:2007xv,Han:2009ss,Cho:2012gd} for other recent methods of mass measurement that do {\em not} use MET for this purpose, similarly to ours above, and Refs.~\cite{Barr:2010hs,Gripaios:2011kk} for a general review of mass measurement methods.
}
%
%
In fact, even if both daughters are visible, a full reconstruction is often afflicted by combinatorial issues. For
example, if the mother particle is 
pair produced and the two of them undergo the same decay process, then it is not clear
on an event-by-event basis which two visible particles to 
use in reconstructing 
%
%
the mother.
Our method clearly avoids this potential problem. 

Note that 
both the above issues come into play in the example considered here, namely, 
top quark pair production.
For example, if the 
$W$ from the top quark decay, in turn, decays leptonically, then the neutrino is invisible.
Of course, in  some cases one can nonetheless determine the neutrino momentum and thus 
reconstruct the $W$. For instance, in semi-leptonic top pair decay we
can impose the neutrino $p_T$
%
%
to be recoiling against the rest of the visible particles 
 and we then enforce the $W$ mass constraint
on the lepton-neutrino combination in order to obtain the neutrino 
longitudinal momentum.
However, 
we face a discrete ambiguity in the second step of the above process.
Moreover, 
even if the $W$ is reconstructed as above, we do not know which $b$-jet to combine it with in order to form the top quark.

It is also clear that our method and the traditional techniques for mass measurement are sensitive to different kind of 
detector effects, for example, our method does not rely on a measurement of the missing transverse momentum/energy in the event, unlike
other methods.
In general, we thus expect 
%
%
that there will be a large degree of complementarity of our method with more traditional ones.

Finally, we emphasize that the proposed technique, despite being based on a fitting function, relies 
only on the minimal assumptions of absence of polarization and the presence of a non-trivial boost distribution
 of the mother particle, i.e., it does not  require any other prior knowledge about the underlying physics model governing the decay of the particle whose mass we want to measure. This suggests that our method will be especially suitable for the mass measurement of new particles 
%
 %
 which might 
 be discovered at the LHC, where  
we (a priori) would not know such details.



We thus envisage
a number of applications of our finding  
about the distributions of the energy of a daughter particle from a two-body decay
(called ``energy-peak" method for short). 
In fact, a number of phenomenological studies along these lines have already appeared.
For example, 
{\em direct} uses of our basic result were made in Ref.~\cite{Chen:2014oha} for measuring the mass of Kaluza-Klein excitation of graviton and also in Refs.~\cite{Kim:2015usa} in the context of a dark matter annihilation model for explaining cosmic ray excesses including the Galactic Center GeV gamma-ray excess.
Meanwhile, there has been a suggestion to use the energy-peak in distinguishing signal from background in {\em searching} for superpartners of the top quark: see Ref.~\cite{Low:2013aza}.
In Ref.~\cite{Agashe:2013eba}, we analyzed
the case of a two-step cascade of 2-body decays of a heavy particle into one invisible and two visible Standard Model ones.
We showed how energy peak measurements can lead to 
the determination of the masses of all three new particles (including the invisible one)
involved in this decay chain; as a more specific example, we studied in detail gluino decay into two bottom quarks and
lightest neutralino (via {\em on}-shell bottom squark) in a $R$-parity conserving supersymmetric model.
Additionally, an application of the 2-body result in distinguishing decays of bottom partners into bottom quark 
accompanied by one vs.~two (massive) invisible particles (as in dark matter models) is worked out in detail in
Ref.~\cite{Agashe:2012fs}. This latter analysis can be generalized to the case of other
three- and two-body decays.
Moreover, further developments
of our 2-body observation are also possible.
As an illustration, an extension to 
3-body decay of a heavy particle into two visible and one invisible ones was considered in \cite{Agashe:2015wwa}, the basic idea being
to slice the 3-body phase-space into multiple 2-body ones (labeled by invariant mass of 2 visible particles). 
Furthermore, an application of this general method to gluino decay into 2 bottom quarks and lightest neutralino via off-shell bottom squark was studied.
On top of these results, 
%
%
a generalization
to the case of a {\em massive} daughter from 2-body decay is in progress \cite{future}.

\noindent
{\bf Note added} \,\, The CMS collaboration
has recently published a measurement of the top quark mass that follows our proposal~\cite{PAS}. The results of this analysis of the 8~TeV LHC dataset, together with preliminary results of the calculation of the missing higher-order contribution~\cite{NLOwork}, indicate very promising prospects for the extraction of the top quark mass with sub-GeV accuracy once more data from the 13 TeV run will be available.  

\noindent
{\bf 2nd Note added} \,\, After our work was submitted, we found that the basic result on which our techniques for mass measurement rely  on had appeared in previous work~\cite{1971NASSP.249.....S} about cosmic ray physics. We remark that our results are more general than those of Ref.~\cite{1971NASSP.249.....S}, which dealt only with the case of scalar decaying particles (i.e., $\pi^0$). In fact, our result here also covers the case of particles with spin. We stress that, to the best of our knowledge, the observation made in Ref.~\cite{1971NASSP.249.....S} 
(and here) had not been applied previously in high-energy particle physics.

\vspace{-0.3cm}

\section*{Acknowledgments}
We would like to thank Alberto Belloni, Roberto Contino, Sarah Eno, Nicholas Hadley, Kirill Melnikov, Sang Eun Lee, Michele Papucci, Raman Sundrum, Jeff Temple, Jesse Thaler, and Kyle Wardlow for comments and discussions, Hongsuk Kang and Young Soo Yoon for help with fitting, and Youngho Yoon for mathematical discussions related to our work. We also thank Roberto Contino for a careful reading of the draft. This work was supported in part by NSF Grant No. PHY-0652363. D.~K. also acknowledges the support from the LHC Theory Initiative graduate fellowship (NSF Grant No. PHY-0969510). The work of R.~F. is also supported by the NSF Grant No. PHY-0910467, and by the Maryland Center for Fundamental Physics. K.~A. and R.~F. acknowledge the hospitality of the Aspen Center for Physics, which is supported by the NSF National Science Foundation Grant No. PHY-1066293.




\begin{thebibliography}{999}



  
\bibitem{Abbiendi:2003rt}
G.~{Abbiendi } {\em et al.}, 
  \href{http://dx.doi.org/10.1140/epjc/s2002-01070-9}{{\em European Physical
  Journal C} {\bfseries 26} 
  321--330},
  \href{http://arxiv.org/abs/arXiv:hep-ex/0203026}{{\ttfamily
  arXiv:hep-ex/0203026}}.


\bibitem{Han:2005mu} 
For a review, see, for example, 
  T.~Han,
  hep-ph/0508097.
%


\bibitem{Alwall:2011fk}
J.~{Alwall} {\em et al.},
  \href{http://dx.doi.org/10.1007/JHEP06(2011)128}{{\em Journal of High Energy
  Physics} {\bfseries 6} 
  128},
  \href{http://arxiv.org/abs/1106.0522}{{\ttfamily arXiv:1106.0522
}}.

\bibitem{Pumplin:2002vw}
J.~Pumplin {\em et al.}, 
 {\em Journal of High Energy
  Physics} {\bfseries 0207} (2002) 012,
\href{http://arxiv.org/abs/hep-ph/0201195}{{\ttfamily arXiv:hep-ph/0201195
}}.

\bibitem{James:2006zz}
F.~James,
``{Statistical methods in experimental physics},''.



\bibitem{Martin:2003sf}
A.~D. {Martin} {\em et al.}
  \href{http://dx.doi.org/10.1140/epjc/s2003-01196-2}{{\em European
  Physical Journal C} {\bfseries 28} 
  455--473},
  \href{http://arxiv.org/abs/arXiv:hep-ph/0211080}{{\ttfamily
  arXiv:hep-ph/0211080}}.

\bibitem{Sjostrand:2006zr}
T.~{Sj{\"o}strand} {\em et al.} 
 \href{http://dx.doi.org/10.1088/1126-6708/2006/05/026}{{\em
  Journal of High Energy Physics} {\bfseries 5} 
   26},
  \href{http://arxiv.org/abs/arXiv:hep-ph/0603175}{{\ttfamily
  arXiv:hep-ph/0603175}}.

\bibitem{Ovyn:2009ys}
S.~{Ovyn} {\em et al.} 
  \href{http://arxiv.org/abs/0903.2225}{{\ttfamily arXiv:0903.2225
}}.

\bibitem{Cacciari:2011rt}
M.~{Cacciari} {\em et al.} 
  \href{http://arxiv.org/abs/1111.6097}{{\ttfamily arXiv:1111.6097
}}.

\bibitem{Cacciari:2006vn}
M.~{Cacciari} and G.~P. {Salam}, 
\href{http://dx.doi.org/10.1016/j.physletb.2006.08.037}{{\em
  Physics Letters B} {\bfseries 641} 
   57--61},
  \href{http://arxiv.org/abs/arXiv:hep-ph/0512210}{{\ttfamily
  arXiv:hep-ph/0512210}}.

\bibitem{Cacciari:2008hb}
M.~{Cacciari} {\em et al.}, 
  \href{http://dx.doi.org/10.1088/1126-6708/2008/04/063}{{\em Journal of High
  Energy Physics} {\bfseries 4}
    63},
  \href{http://arxiv.org/abs/0802.1189}{{\ttfamily arXiv:0802.1189
}}.

\bibitem{ATLAS-CONF-2012-097}
{ATLAS Collaboration}, 
  ATLAS-CONF-2012-097 


\bibitem{Lester:1999et}
C.~G. {Lester} and D.~J. {Summers}, ``{Measuring masses of semi-invisibly
  decaying particle pairs produced at hadron colliders},''
  \href{http://dx.doi.org/10.1016/S0370-2693(99)00945-4}{{\em Physics Letters
  B} {\bfseries 463} (Sept., 1999) 99--103},
  \href{http://arxiv.org/abs/arXiv:hep-ph/9906349}{{\ttfamily
  arXiv:hep-ph/9906349}}.
  

\bibitem{Barr:2003fj}
A.~{Barr}, C.~{Lester}, and P.~{Stephens}, ``{A variable for measuring masses
  at hadron colliders when missing energy is expected $m_{T2}$: the truth
  behind the glamour},''
  \href{http://dx.doi.org/10.1088/0954-3899/29/10/304}{{\em Journal of Physics
  G Nuclear Physics} {\bfseries 29} (Oct., 2003) 2343--2363},
  \href{http://arxiv.org/abs/arXiv:hep-ph/0304226}{{\ttfamily
  arXiv:hep-ph/0304226}}.

\bibitem{Cho:2007qv} 
  W.~S.~Cho, K.~Choi, Y.~G.~Kim and C.~B.~Park,
  ``Gluino Stransverse Mass,''
    \href{http://dx.doi.org/10.1103/PhysRevLett.100.171801}{ Phys.\ Rev.\ Lett.\  {\bf 100}, 171801 (2008)},
    \href{http://arxiv.org/abs/0709.0288}{{\ttfamily arXiv:0709.0288 [hep-ph]}};
%
``{Measuring
  superparticle masses at hadron collider using the transverse mass kink},''
  \href{http://dx.doi.org/10.1088/1126-6708/2008/02/035}{{\em Journal of High
  Energy Physics} {\bfseries 2} (Feb., 2008) 35},
  \href{http://arxiv.org/abs/0711.4526}{{\ttfamily arXiv:0711.4526 [hep-ph]}}; 
A.~J. {Barr}, B.~{Gripaios}, and C.~G. {Lester}, ``{Weighing wimps with kinks
  at colliders: invisible particle mass measurements from endpoints},'' \href{http://dx.doi.org/10.1088/1126-6708/2008/02/014}{{\em Journal of High
  Energy Physics} {\bfseries 2} (Feb., 2008) 14},
 \href{http://arxiv.org/abs/0711.4008}{{\ttfamily arXiv:0711.4008 [hep-ph]}}.


\bibitem{Barr:2011ao} 
For a guide to the literature on transverse mass variables, see, for example, A.~J.~Barr, T.~J.~Khoo, P.~Konar, K.~Kong, C.~G.~Lester, K.~T.~Matchev and M.~Park,
  ``Guide to transverse projections and mass-constraining variables,''
  \href{http://dx.doi.org/10.1103/PhysRevD.84.095031}{Phys.\ Rev.\ D {\bf 84}, 095031 (2011)}
   \href{http://arxiv.org/abs/arXiv:1105.2977}{{\ttfamily
  arXiv:1105.2977 [hep-ph]}}.
  %
A very recent proposal is in  
A.~Ismail, R.~Schwienhorst, J.~S.~Virzi and D.~G.~E.~Walker,
  Phys.\ Rev.\ D {\bf 91}, no. 7, 074002 (2015)
  [arXiv:1409.2868 [hep-ph]].

\bibitem{Cho:2014naa} 
  W.~S.~Cho, J.~S.~Gainer, D.~Kim, K.~T.~Matchev, F.~Moortgat, L.~Pape and M.~Park,
  ``On-shell constrained $M_2$ variables with applications to mass measurements and topology disambiguation,''
  JHEP {\bf 1408}, 070 (2014)
  [arXiv:1401.1449 [hep-ph]].


\bibitem{Randall:2008rw} 
  L.~Randall and D.~Tucker-Smith,
  ``Dijet Searches for Supersymmetry at the LHC,''
  
    \href{http://dx.doi.org/10.1103/PhysRevLett.101.221803}{Phys.\ Rev.\ Lett.\  {\bf 101}, 221803 (2008)}
    \href{http://arxiv.org/abs/arXiv:0806.1049}{{\ttfamily
  arXiv:0806.1049 [hep-ph]}}
 

  
\bibitem{Rogan:2010kb} 
  C.~Rogan,
  ``Kinematical variables towards new dynamics at the LHC,''
  arXiv:1006.2727 [hep-ph];
  M.~R.~Buckley, J.~D.~Lykken, C.~Rogan and M.~Spiropulu,
  Phys.\ Rev.\ D {\bf 89}, no. 5, 055020 (2014)
  [arXiv:1310.4827 [hep-ph]].


\bibitem{Gripaios:2011kk}
B.~{Gripaios}, ``{Tools for Extracting New Physics in Events with Missing
  Transverse Momentum},''
  \href{http://dx.doi.org/10.1142/S0217751X11054826}{{\em International Journal
  of Modern Physics A} {\bfseries 26} (2011) 4881--4900},
  \href{http://arxiv.org/abs/1110.4502}{{\ttfamily arXiv:1110.4502 [hep-ph]}}.



\bibitem{Cheng:2007xv}
H.-C. {Cheng}, J.~F. {Gunion}, Z.~{Han}, G.~{Marandella}, and B.~{McElrath},
  ``{Mass determination in SUSY-like events with missing energy},''
  \href{http://dx.doi.org/10.1088/1126-6708/2007/12/076}{{\em Journal of High
  Energy Physics} {\bfseries 12} (Dec., 2007) 76},
  \href{http://arxiv.org/abs/0707.0030}{{\ttfamily arXiv:0707.0030 [hep-ph]}}.
%
H.-C. {Cheng}, D.~{Engelhardt}, J.~F. {Gunion}, Z.~{Han}, and B.~{McElrath},
  ``{Accurate Mass Determinations in Decay Chains with Missing Energy},''
  \href{http://dx.doi.org/10.1103/PhysRevLett.100.252001}{{\em Physical Review
  Letters} {\bfseries 100} no.~25, (June, 2008) 252001},
  \href{http://arxiv.org/abs/0802.4290}{{\ttfamily arXiv:0802.4290 [hep-ph]}}.
 %
H.-C. {Cheng}, J.~F. {Gunion}, Z.~{Han}, and B.~{McElrath}, ``{Accurate mass
  determinations in decay chains with missing energy: II},''
  \href{http://dx.doi.org/10.1103/PhysRevD.80.035020}{{\em Phys. Rev. D}
  {\bfseries 80} no.~3, (Aug., 2009) 035020},
  \href{http://arxiv.org/abs/0905.1344}{{\ttfamily arXiv:0905.1344 [hep-ph]}}.


  
\bibitem{Han:2009ss} 
T.~{Han}, I.-W. {Kim}, and J.~{Song}, ``{Kinematic cusps: Determining the
  missing particle mass at colliders},''
  \href{http://dx.doi.org/10.1016/j.physletb.2010.09.010}{{\em Physics Letters
  B} {\bfseries 693} (Oct., 2010) 575--579},
  \href{http://arxiv.org/abs/0906.5009}{{\ttfamily arXiv:0906.5009 [hep-ph]}}; ``{Kinematic Cusps With Two Missing
  Particles I: Antler Decay Topology},'' {\em ArXiv e-prints} (June, 2012) ,
  \href{http://arxiv.org/abs/1206.5633}{{\ttfamily arXiv:1206.5633 [hep-ph]}}; and
``{Kinematic Cusps with Two Missing
  Particles II: Cascade Decay Topology},''   Phys.\ Rev.\ D {\bf 87}, no. 3, 035004 (2013), {\em ArXiv e-prints} (June, 2012) ,
  \href{http://arxiv.org/abs/1206.5641}{{\ttfamily arXiv:1206.5641 [hep-ph]}}.


 \bibitem{Cho:2012gd}
W.~S. {Cho}, D.~{Kim}, K.~T. {Matchev}, and M.~{Park}, ``{Cracking the dark
  matter code at the LHC},'' {\em ArXiv e-prints} (June, 2012) ,
  \href{http://arxiv.org/abs/1206.1546}{{\ttfamily arXiv:1206.1546 [hep-ph]}}.
  
  \bibitem{Barr:2010hs}
A.~J. {Barr} and C.~G. {Lester}, ``{A review of the mass measurement techniques
  proposed for the Large Hadron Collider},''
  \href{http://dx.doi.org/10.1088/0954-3899/37/12/123001}{{\em Journal of
  Physics G Nuclear Physics} {\bfseries 37} no.~12, (Dec., 2010) 123001},
  \href{http://arxiv.org/abs/1004.2732}{{\ttfamily arXiv:1004.2732 [hep-ph]}}.



\bibitem{Chen:2014oha} 
  C.~Y.~Chen, H.~Davoudiasl and D.~Kim,
  ``Z with missing energy as a warped graviton signal at hadron colliders,''
  Phys.\ Rev.\ D {\bf 89}, no. 9, 096007 (2014)
  [arXiv:1403.3399 [hep-ph]].
  
  
  
  
\bibitem{Kim:2015usa} 
  D.~Kim and J.~C.~Park,
  ``Energy peak: back to the Galactic Center GeV gamma-ray excess,''
  arXiv:1507.07922 [hep-ph] and 
  %
  ``An alternative interpretation for cosmic ray peaks,''
  arXiv:1508.06640 [hep-ph].


\bibitem{Agashe:2013eba} 
  K.~Agashe, R.~Franceschini and D.~Kim,
  ``Using Energy Peaks to Measure New Particle Masses,''
  JHEP {\bf 1411}, 059 (2014)
  [arXiv:1309.4776 [hep-ph]].


\bibitem{Agashe:2015wwa} 
  K.~Agashe, R.~Franceschini, D.~Kim and K.~Wardlow,
  ``Mass Measurement Using Energy Spectra in Three-body Decays,''
  arXiv:1503.03836 [hep-ph].



\bibitem{Low:2013aza} 
  I.~Low,
  ``Polarized Charginos (and Tops) in Stop Decays,''
  arXiv:1304.0491 [hep-ph].
  %


\bibitem{Agashe:2012fs} 
  K.~Agashe, R.~Franceschini, D.~Kim and K.~Wardlow,
  ``Using Energy Peaks to Count Dark Matter Particles in Decays,''
  arXiv:1212.5230 [hep-ph], 
  
  
\bibitem{future}
K.~Agashe,  R.~Franceschini, D.~Kim and S.~Hong (in preparation), ``Energy-peak for a massive daughter from a two-body
decay".
  
 \bibitem{PAS}
{CMS Collaboration}
%
{``Measurement of the top-quark mass from the b jet energy spectrum"}, 
%
CMS PAS TOP-15-002.

\bibitem{NLOwork}
K.~Agashe, R.~Franceschini, D.~Kim, M.~Schulze, {``NLO precise top quark mass from the bottom energy peak"}  (in preparation).

 \bibitem{1971NASSP.249.....S}
F.~W. {Stecker}, ``{Cosmic gamma rays},'' {\em NASA Special Publication}
  {\bfseries 249} (1971).
 



    
\end{thebibliography}
\end{document}